\documentclass[reprint,amsmath,amssymb,aps,pra,showpacs]{revtex4-1}

\pdfoutput=1

\usepackage{graphicx}
\usepackage{dcolumn}
\usepackage{bm}
\usepackage{hyperref}
\usepackage[caption=false]{subfig}

\hypersetup{
	colorlinks = true,
	urlcolor   = blue,
	linkcolor  = blue,
	citecolor  = blue
}

\renewcommand*{\phi}{\varphi}
\renewcommand*{\epsilon}{\varepsilon}
\renewcommand*{\le}{\leqslant}
\renewcommand*{\ge}{\geqslant}
\renewcommand*{\Re}{\mathop{\mathrm{Re}}\nolimits}
\renewcommand*{\Im}{\mathop{\mathrm{Im}}\nolimits}

\DeclareMathOperator{\Tr}{Tr}
\DeclareMathOperator{\rank}{rank}
\DeclareMathOperator{\diag}{diag}

\newcommand*{\bra}[1]{\langle #1 |}
\newcommand*{\ket}[1]{| #1 \rangle}
\newcommand*{\scalprod}[2]{\langle #1 | #2 \rangle}

\newcommand*{\dyad}[1]{| #1 \rangle \langle #1 |}

\newcommand*{\HG}{\text{HG}}
\newcommand*{\e}[1]{\mathrm{e}{#1}}

\newtheorem{condition}{Condition}

\newcolumntype{2}{D{.}{.}{2}}
\newcolumntype{3}{D{.}{.}{3}}
\newcolumntype{4}{D{.}{.}{4}}
\newcolumntype{5}{D{.}{.}{5}}
\newcolumntype{6}{D{.}{.}{6}}

\begin{document}

\title{Adaptive quantum tomography of high-dimensional bipartite systems}

\author{G.\,I.\,Struchalin}
	\email{struchalin.gleb@physics.msu.ru}
\author{E.\,V.\,Kovlakov}
\author{S.\,S.\,Straupe}
\author{S.\,P.\,Kulik}
\affiliation{Faculty of Physics, M.\,V.\,Lomonosov Moscow State University}
\affiliation{Quantum Technologies Centre,  M.\,V.\,Lomonosov Moscow State University, 119991, Moscow, Russia}

\date{\today}

\begin{abstract}

Adaptive measurements have recently been shown to significantly improve the performance of quantum state and process tomography. However, the existing methods either cannot be straightforwardly applied to high-dimensional systems or are prohibitively computationally expensive. Here we propose and experimentally implement a novel tomographic protocol specially designed for the reconstruction of high-dimensional quantum states. The protocol shows qualitative improvement in infidelity scaling with the number of measurements and is fast enough to allow for complete state tomography of states with dimensionality up to 36. 

\end{abstract}


\maketitle

\section{Introduction\label{sec:Introduction}}

Quantum state tomography is a procedure which allows one to reconstruct the full density matrix of a quantum state from the outcomes of measurements on an ensemble of systems prepared in that state \cite{Paris_Book2004}. Similarly, quantum process tomography reconstructs the $\chi$-matrix, describing a transformation of a quantum system in a most general form \cite{Nielsen_JMO97}. In the era of rapidly developing quantum technologies, quantum tomography becomes one of the critical primitives, allowing for experimental analysis and debugging of quantum devices under development. It is therefore crucial to develop tomographic protocols capable of reconstruction of complex high-dimensional states and processes. Although, full state tomography of a system living in a $D$-dimensional Hilbert space requires at least $D^2$ different measurements and is, therefore, not a scalable procedure, one still needs protocols, which are tractable for at least few-qubit states.

Precision of the tomographic estimate significantly depends on the choice of the protocol -- i.e. the specific set of measurements performed. The quality of reconstruction may be significantly improved, if these measurements are chosen adaptively, relying on the previous data to tune the following measurements to increase the statistical significance of the observed outcomes. Although, first ideas and implementations following this line of thought appeared quite early \cite{Massar_PRA00,Freyberger_PRA00}, adaptive methods in quantum tomography have recently seen significant advances (see \cite{Straupe_JETP2016} for a review). One approach is to use Bayesian methods of optimal experimental design \cite{Houlsby_PRA12}, which was experimentally realized for single-qubit \cite{Kravtsov_PRA13} and two-qubit states \cite{Kulik_PRA16}, as well as single-qubit quantum processes \cite{Kulik_PRA17}. This method is completely general and theoretically attractive, however involved computational resources are so high, that it becomes impractical for high-dimensions. 

Another approach was suggested in \cite{Massar_PRA00} and later in \cite{Rehacek04}, and realized experimentally for qubits in \cite{Steinberg_PRL13} and \cite{Guo_NPJQI16}. It essentially suggests to perform state estimation as a two-step process, first obtain an estimate by measurements in an arbitrary basis, and then change the measurement basis to the eigenbasis of the estimated density matrix. The problem with straightforward generalization of this approach to high-dimensions is the fact, that the eigenstates of the estimate will almost certainly be entangled high-dimensional states, and realization of the corresponding projective measurements experimentally is usually extremely challenging. There are other ideas and approaches to designing optimal adaptive strategies for quantum tomography \cite{Murao_PRA12,Takeuchi_PRL12,Hen_NJP15}, however none of them has experimentally gone beyond two qubits \cite{Stefanov_OptLett14,Guo_npjQI2017,wang_SciChina2016}. 

A rare exception among adaptive protocols is the recently suggested self-guided tomography \cite{Ferrie_PRL14}, which was shown to be tractable for at least 7 qubits in numerical simulations. This protocol is however directly applicable only for the reconstruction of pure states, and should be extended with Bayesian data processing to allow for mixed states reconstruction \cite{Granade_NJP2017}. In this case it shares the same computational difficulties with other Bayesian protocols. Here again, experimental implementations were only limited to two-qubit states \cite{Ferrie_PRL2016}.

In this article we present a novel adaptive protocol, which is specially tailored for high-dimensional bipartite states. Such states are ubiquitous in many experimental settings, for example, in experiments with orbital-angular momentum and entangled spatial states of photons \cite{Torres_NaturePhys2007}. Our protocol utilizes only factorized measurements performed separately on the subsystems, which makes it practical for implementation. It is also independent of the choice of a statistical estimation procedure. We provide intuitive arguments explaining the reasons of the increased estimation accuracy and confirm them by numerical simulations and real experiments. The experimental testbed for the protocol is the reconstruction of high-dimensional (up to $D=36$) entangled spatial states of photon pairs.



\section{Algorithm\label{sec:Algorithm}}

\subsection{Protocol accuracy\label{sec:FidelityDistribution}}
An estimator~$\hat \rho$ is a map from random measurement outcomes to the system Hilbert space $\mathcal H_D$ of dimension~$D$. Therefore, an estimator itself and all quantities involving it are random variables. Our protocol was inspired by the existing theory of universal statistical distribution for fidelity $F(\rho, \hat \rho) = \Tr^2 \sqrt{\rho^{1/2} \hat \rho \rho^{1/2}}$ between the true state~$\rho$ and the estimator~$\hat \rho$~\cite{Bogdanov_JETP2009}. Let us outline the main results of this theory for convenience.

The theory is valid for a maximum-likelihood estimator and provides an asymptotic distribution of fidelity $F(\rho, \hat \rho)$ in the limit of infinitely large number of detected events~$N$, whenever a measurement protocol is given. Measurements are characterized by a positive operator-valued measures (POVMs) $\{\mathcal M_\alpha\}$. POVM elements $M_{\alpha \gamma} \in \mathcal M_\alpha$ satisfy the normalization constraint: $\sum_{\gamma} M_{\alpha \gamma} = 1_D$, where~$1_D$ is a $D$-dimensional identity matrix. We will be interested in projective measurements in some basis, i.\,e. each POVM consists of $D$ rank-1 projectors: $\mathcal M_\alpha = \{\ket{\phi_{\alpha \gamma}} \bra{\phi_{\alpha \gamma}}\}_{\gamma = 1}^{D}$. The probability of obtaining an outcome~$\gamma$ in a measurement $\alpha$ with the system being in the state~$\rho$ is given by Born's rule:
\begin{equation}
p_{\alpha \gamma} = \Tr(M_{\alpha \gamma} \rho). \label{eq:BornRule}
\end{equation}

Suppose~$\rho$ is a rank-$R$ state with only~$R$ nonzero eigenvalues~$\lambda_k$. It can be purified in the extended Hilbert space $\mathcal H_{RD} = \mathcal H_R \otimes \mathcal H_D$ of dimension $R \times D$: $\rho = \Tr_{R} \ket{\Psi} \bra{\Psi}$. The purification~$\ket{\Psi}$ is arbitrary up to unitary transformations on the auxiliary system~$\mathcal H_R$, one possible choice is
\begin{equation}
\ket{\Psi} = \sum_{k=1}^{R} \sqrt{\lambda_k} \ket{k} \otimes \ket{\psi_k}, \label{eq:Purification}
\end{equation}
where~$\ket{\psi_k}$ are the eigenvectors of~$\rho$ corresponding to nonzero eigenvalues~$\lambda_k$ (henceforward, $\lambda_k$ are assumed to be sorted in decreasing order). Let $M_{\alpha \gamma}' = 1_R \otimes M_{\alpha \gamma}$ be an ``extended'' measurement operator acting on $\mathcal H_{RD}$ space, then the probability in~(\ref{eq:BornRule}) is invariant under the replacement $\rho \rightarrow \ket{\Psi}\bra{\Psi}$ and $M_{\alpha \gamma} \rightarrow M_{\alpha \gamma}'$.

As a final preparatory step let us switch from complex to real-valued vectors and matrices. Indeed, every complex matrix~$A$ and column-vector~$v$ can be viewed as a real-valued matrix~$A_\text{re}$ and a vector~$v_\text{re}$ of doubled dimension:
\begin{equation}
A_\text{re} = \begin{bmatrix}
\Re A & -\Im A \\
\Im A & \Re A
\end{bmatrix}, \quad
v_\text{re} = \begin{bmatrix}
\Re v \\
\Im v
\end{bmatrix}. \label{eq:ComplexRealIsomorphism}
\end{equation}
Linear algebraic expressions, e.\,g. $w = Av$, maintain their form under this isomorphism: $w_\text{re} = A_\text{re} v_\text{re}$. The hermitian conjugation operation is replaced by transposition alone: $A^\dagger \rightarrow A_\text{re}^T$. Using the purification~(\ref{eq:Purification}) and isomorphism~(\ref{eq:ComplexRealIsomorphism}), Born's rule~(\ref{eq:BornRule}) can be rewritten as
\begin{equation}
p_{\alpha \gamma} = c^T O_{\alpha \gamma} c,
\end{equation}
where $c = \ket{\Psi}_\text{re}$ and $O_{\alpha \gamma} = (M_{\alpha \gamma}')_\text{re}$.

The uncertainty of an asymptotically efficient estimator is characterized by the Fisher information matrix~$H$ (via its inverse):
\begin{equation}
H_{ij} = \left \langle
\frac{\partial \ln \mathcal L(c; \{n_{\alpha \gamma}\})}{\partial c_i}
\frac{\partial \ln \mathcal L(c; \{n_{\alpha \gamma}\})}{\partial c_j}
\right \rangle, \label{eq:FisherInformation}
\end{equation}
where $\mathcal L(c; \{n_{\alpha \gamma}\})$ is a likelihood function, and expectation is carried out over different measurement outcomes~$\{n_{\alpha \gamma}\}$. The Fisher information matrix~$H$ is a symmetric real-valued matrix of size $2RD \times 2RD$. If a tomographic protocol is informationally complete, $H$ has $2RD - R^2$ strictly positive singular values~$\sigma_i$, while other $R^2$ ones are exactly zero. Henceforth we will assume that~$\sigma_i$ are sorted in decreasing order.

Fidelity~$F(\rho, \hat \rho)$ between the true state~$\rho$ and an asymptotically efficient, e.\,g. a maximum-likelihood, estimator~$\hat \rho$ is closely related to the singular values~$\sigma_i$ of~$H$. An asymptotic distribution of~$F(\rho, \hat \rho)$ in the limit of infinitely many observations can be represented as follows \cite{Bogdanov_JETP2009,Kulik_PRL10}:
\begin{equation}
1 - F = \sum_{i = 2}^{\nu + 1} \frac{1}{\sigma_i} \xi_i^2, \label{eq:FidelityDistribution}
\end{equation}
where $\nu = 2RD - R^2 - 1$, and $\xi_i \sim \mathcal N(0, 1)$ are identically and independently distributed normal random variables with zero mean and unit variance. The sum~(\ref{eq:FidelityDistribution}) contains~$\nu$ terms, which is equal to the number of degrees of freedom for the rank-$R$ quantum state, for example $\nu = 2D-2$ for pure states, and $\nu = D^2 - 1$ for full rank states. The right-hand side of~(\ref{eq:FidelityDistribution}) is a sort of generalized chi-squared distribution, useful series representations of its distribution function can be found in~\cite{Canada_CMA1984, Provost_Revstat2013}. Expectation $\langle 1 - F \rangle$ and standard deviation $\Delta (1 - F)$ of $1-F$ are obtained straightforwardly:
\begin{equation}
\langle 1 - F \rangle = \sum_{i = 2}^{\nu + 1} \frac{1}{\sigma_i}, \quad
\Delta (1 - F) = \sqrt{\sum_{i = 2}^{\nu + 1} \frac{2}{\sigma_i^2}}. \label{eq:FidelityMeanStdev}
\end{equation}

In the following we will be interested in the likelihood function~$\mathcal L(c)$, expressed as a product of Poissonian probabilities, since it is usually the case in experiments with photon counting:
\begin{equation}
\mathcal L(c) = \prod_{\alpha \gamma} \frac{[ p_{\alpha \gamma}(c) b_{\alpha \gamma} ]^{n_{\alpha \gamma}}}{n_{\alpha \gamma}!} e^{-p_{\alpha \gamma}(c) b_{\alpha \gamma}}, \label{eq:PoissonLikelihood}
\end{equation}
here $b_{\alpha \gamma}$ are constants proportional to exposition time, $n_{\alpha \gamma}$ are the numbers of detected counts, $\sum_{\alpha \gamma} n_{\alpha \gamma} = N$. Note, that~(\ref{eq:PoissonLikelihood}) actually covers the canonical for quantum tomography case of \emph{multinomial} likelihood, $\mathcal L_\text{mult} \propto \prod_{\alpha \gamma} p_{\alpha \gamma}^{n_{\alpha \gamma}}$, when $b_{\alpha \gamma}$ does not vary with index~$\gamma$: $b_{\alpha \gamma} = b_\alpha$. In this case, $\sum_{\alpha \gamma} p_{\alpha \gamma} b_{\alpha \gamma} = \sum_{\alpha} b_\alpha$ does not depend on~$\rho$, and the exponent can be absorbed by the proportionality sign.

Fisher information~$H$ for the Poissonian likelihood~(\ref{eq:PoissonLikelihood}) is,
\begin{equation}
H = \sum_{\alpha \gamma} \frac{4 b_{\alpha \gamma}}{p_{\alpha \gamma}} O_{\alpha \gamma} c c^T O_{\alpha \gamma}. \label{eq:FisherInformationPoisson}
\end{equation}
Given that total number of counts~$N$ is fixed, the equality  $c^T H c = 4N$ holds, since $p_{\alpha \gamma} b_{\alpha \gamma}$ is equal to the expectation $\langle n_{\alpha \gamma} \rangle$. The largest singular value, $\sigma_1 = 4N$, corresponds to the vector~$c$. Other nonzero singular values also grow as fast as~$N$ in the asymptotic limit: $\sigma_i \propto N$. Thus both the expectation and the standard deviation~(\ref{eq:FidelityMeanStdev}) are inversely proportional to the total number of counts detected~$N$: $\langle 1 - F \rangle \propto 1/N$ and $\Delta (1 - F)  \propto 1/N$.

It is implicitly assumed in the derivation of~(\ref{eq:FidelityDistribution}), that the rank~$R_s$ of the true state~$\rho$ matches the rank~$R_e$ of the estimator~$\hat \rho$ (i.\,e. the likelihood is optimized over the set of states of rank~$R_e$): $R_s = R_e = R$. In real tomographic experiments due to instrumental imperfections, nominally, the true state is always full rank, $R_s = D$, however some of the eigenvalues may be relatively small. Estimation of the state mixedness compels an experimenter to reconstruct the state as a full-rank one, $R_e = D$. Formally, there is no problem at all, because~$R_s = R_e$, but the presence of tiny eigenvalues reduces the estimation accuracy~$1-F(\rho, \hat \rho)$ dramatically~-- the true state behaves effectively as a rank deficient one. In general, loss of accuracy occurs, when $R_s < R_e$ (there is no need for~$R_e$ being equal to~$D$). It is well known that fidelity in this case can degrade up to $\langle 1 - F \rangle \propto 1/\sqrt{N}$~\cite{Bagan_PRL2006}.

The reason asymptotic $1/N$ does not hold anymore is that some of the singular values~$\sigma_i$ in the sum~(\ref{eq:FidelityDistribution}) become zero. The number of terms in~(\ref{eq:FidelityDistribution}) is equal to $\nu_e + 1 \equiv 2 R_e D - R_e^2$ and is determined by the estimator rank~$R_e$. On the other hand, the number of nonzero singular values or the rank of the Fisher information matrix~$H$ is related to the true state rank~$R_s$: $\rank H = \nu_s + 1 \equiv 2 R_s D - R_s^2$.

Let us consider an example: suppose a rank-3 state with its nonzero eigenvalues being $\lambda_1, \lambda_2, \lambda_ 3$ is measured using some protocol $\{M_{\alpha \gamma}\}$. One can calculate the Fisher information $H_3(\lambda_1, \lambda_2, \lambda_3)$, assuming the estimator rank is $R_e = 3$. Obviously, $\rank H_3(\lambda_1, \lambda_2, \lambda_3) = \nu(R = 3) + 1$. Now we take the limit $\lambda_3 \to 0$, obtaining $H_3(\lambda_1, \lambda_2, 0)$, which corresponds to the case $R_s = 2$ and $R_e = 3$. Our goal is to find $\rank H_3(\lambda_1, \lambda_2, 0)$. To make things more transparent we consider the term $O_{\alpha \gamma} c$ in~(\ref{eq:FisherInformationPoisson}) or, equivalently, $M'_{\alpha \gamma} \ket{\Psi}$ due to the isomorphism~(\ref{eq:ComplexRealIsomorphism}). $M'_{\alpha \gamma} \ket{\Psi}$ has the following block structure:
\begin{equation}
\begin{bmatrix}
M_{\alpha \gamma} & 0 & 0 \\
0 & M_{\alpha \gamma} & 0 \\
0 & 0 & M_{\alpha \gamma} \\
\end{bmatrix}
\begin{bmatrix}
\sqrt{\lambda_1} \ket{\psi_1} \\
\sqrt{\lambda_2} \ket{\psi_2} \\
0
\end{bmatrix} = 
\begin{bmatrix}
\sqrt{\lambda_1} M_{\alpha \gamma} \ket{\psi_1} \\
\sqrt{\lambda_2} M_{\alpha \gamma} \ket{\psi_2} \\
0
\end{bmatrix}
\end{equation}
Now, if one calculates the matrices $O_{\alpha \gamma} c c^T O_{\alpha \gamma}$ to obtain the Fisher information $H_3(\lambda_1, \lambda_2, 0)$, then the specific rows and columns are exactly zero. This is valid if $p_{\alpha \gamma} \ne 0$ for all operators~$M_{\alpha \gamma}$, which usually happens for static measurement protocols. $H_3(\lambda_1, \lambda_2, 0)$ has the form of a Fisher information matrix $H_2(\lambda_1, \lambda_2)$, computed for $R_e = R_s = 2$ with some zero-valued rows and columns inserted. Obviously, $\rank H_3(\lambda_1, \lambda_2, 0) = \rank H_2(\lambda_1, \lambda_2) = \nu(R = 2) + 1$. Therefore $\sigma_i = 0$ for $i = \nu(R = 2) + 2, \dots, \nu(R = 3) + 1$ in~(\ref{eq:FidelityDistribution}).

\subsection{Estimator-orthogonal measurements}
Measurements for which $p_{\alpha \gamma} \approx 0$ are of special interest. Even though $O_{\alpha \gamma} c c^T O_{\alpha \gamma}$ contains rows and columns with nearly vanishing elements, when $\lambda_3 \approx 0$, they can be magnified by the factor $1/p_{\alpha \gamma} \gg 1$. If sufficient amount of measurements obey $p_{\alpha \gamma} \approx 0$, then the matrix~$H_3$ has no tiny singular values~$\sigma_i$ for $i = 2, \dots \nu_e + 1$, required in~(\ref{eq:FidelityDistribution}), and accuracy of tomography is high. In the limit $\lambda_3 \to 0$ the measurement operators~$M_{\alpha \gamma}$ should be chosen in such a way, that a strict equality $p_{\alpha \gamma} = 0$ holds, to preserve a convergence rate~$1/N$.

One may hope that the protocol, which maintains $1/N$~convergence in the extreme situation $R_s < R_e$, will also have superior accuracy in the situation of small (but nonzero) eigenvalues~$\lambda_i$ of the true state. Therefore, the case $R_s < R_e$ is considered further. We call a measurement~$M_{\alpha \gamma}$ \emph{orthogonal} to a projector~$\dyad{\psi}$ if $\Tr (M_{\alpha \gamma} \dyad{\psi}) = 0$. This implies $M_{\alpha \gamma} \ket{\psi} = 0$ and vice versa due to positivity of $M_{\alpha \gamma}$. Clearly, the aforementioned example can be transferred in full analogy to different combinations of $R_s < R_e \le D$. Now we are ready to formulate the necessary condition for a protocol to maintain convergence~$1/N$ in the presence of discrepancy between the true state rank~$R_s$ and the estimator rank~$R_e$, $R_s < R_e$:
\begin{condition}[necessary]
	\label{cond:Necessary}
	The protocol must contain a measurement~$M_{\alpha \gamma}$ which is orthogonal to the projectors on the eigenvectors~$\ket{\psi_k}$ corresponding to the nonzero eigenvalues of the true state: $M_{\alpha \gamma} \ket{\psi_k} = 0, k = 1, \dots, R_s$.
\end{condition}
This condition means that the measurement~$M_{\alpha \gamma}$ has zero outcome probability: $p_{\alpha \gamma} = 0$. Of course, if an informationally complete protocol contains only one orthogonal operator~$M_{\alpha \gamma}$ then it is not sufficient to improve convergence. The rank of the Fisher information matrix~$H$ is limited by $\nu_s + 1$ if there are no orthogonal measurements. Each independent orthogonal measurement increments the rank by one above this value, until the maximum rank $\nu_e + 1$ is not reached. Therefore, the following sufficient condition holds:
\begin{condition}[sufficient]
	\label{cond:Sufficient}
	The protocol should contain $\nu_e - \nu_s = (R_e - R_s)(2D - R_e - R_s)$ independent measurements~$M_{\alpha \gamma}$, satisfying the condition~\ref{cond:Necessary}.
\end{condition}

It seems that orthogonal measurements demand exact knowledge of the true state~$\rho$, but as in other adaptive protocols with measurement basis alignment \cite{Rehacek04,Steinberg_PRL13}, the estimator-orthogonal protocol aligns the measurements according to the current estimator~$\hat \rho$. The true state eigenvectors are replaced by the estimator eigenvectors. The rank~$R_s$ of the true state is usually also unknown in advance (otherwise one can equate the estimator rank~$R_e$ with the state rank~$R_s$) and, hence, in general we suggest to tune the protocol for all ranks of the input state~-- one should find the measurements orthogonal to $K = 1, \dots, R_e - 1$ eigenvectors subsequently. Firstly, let $K = 1$ and measurements are found to be orthogonal to the eigenvector with the largest eigenvalue (tune for rank-1 states), then set $K = 2$ and subsequent measurements are orthogonalized with respect to the first two eigenvectors (tune for rank-2 states), etc. The protocol, obtained in such a way, has an optimal convergence $1/N$ regardless of the true state rank. However,  the possible values of~$K$ may be specified by some a priori knowledge if available.

\subsection{Factorized measurements}
A high-dimensional quantum system usually has a natural separation into subsystems (tensor product structure), and measurements performed on its parts separately, which we will call \emph{factorized} measurements, are much easier to implement in experiment then \emph{general} measurements on the whole system. According to the condition~\ref{cond:Necessary}, measurements should be orthogonal to the eigenvectors~$\ket{\psi_k}$ of the true state~$\rho$ (or the current estimator~$\hat \rho$), which are almost certainly entangled. But the restriction to factorized measurements poses additional constraints, and a natural question arises: do factorized \emph{and} estimator-orthogonal measurements exist? A short answer is: they do exist if the number~$K$ of vectors to orthogonalize to does not exceed a certain limit~$K_\text{max}$. This means that the accuracy of the estimator-orthogonal protocol with factorized measurements will degrade for the states with rank $R_s > K_\text{max}$.

In a simplest case of a bipartite system and a pure true state, a Schmidt decomposition can be used to find estimator-orthogonal measurements. Indeed, there is only one eigenvector~$\ket{\psi_1}$ with a nonzero eigenvalue ($K = 1$). Its Schmidt decomposition is $\ket{\psi_1} = \sum_i \sqrt{\mu_i} \ket{i} \otimes \ket{i}$, where~$\mu_i$ are the eigenvalues of the reduced density operator. Obviously, factorized vectors $\ket{i} \otimes \ket{j}$, $i \ne j$, are orthogonal to $\ket{\psi_1}$~\footnote{Actually, there is an infinite number of different factorized orthogonal vectors, which do not have this trivial form}. The desired measurements~$M_{\alpha \gamma}$ are the projectors onto these vectors: $M_{\alpha \gamma} = \ket{i} \bra{i} \otimes \ket{j} \bra{j}$.

In general, the existence of measurements~$M^{(K)}$, which are factorized and orthogonal to~$K$ entangled vectors is closely related to the maximal dimension of a \emph{completely entangled} subspace~\cite{Parthasarathy_PIAS2004}. Suppose a Hilbert space~$\mathcal H_D$ of a $D$-dimensional system consists of~$l$ components: $\mathcal H_D = \mathcal H_{d_1} \otimes \mathcal H_{d_2} \otimes \dots \otimes \mathcal H_{d_l}$, where dimensions~$d_i$ of components~$\mathcal H_{d_i}$ obey $d_1 d_2 \dots d_l = D$. A subspace~$S_E \subset \mathcal H_D$ is said to be completely entangled if it contains no factorized vectors. The maximal possible dimension of a completely entangled subspace is
\begin{equation}
D_E \equiv \max_{S_E \in \mathcal E} \dim S_E = D - (d_1 + \dots + d_l) + l - 1, \label{eq:DimMaxEntSubspace}
\end{equation}
where~$\mathcal E$ is the set of all completely entangled subspaces.

Let $S_\psi \subset \mathcal H_D$ be a subspace spanned by~$K$ vectors~$\ket{\psi_k}$ ($\dim S_\psi = K$), and $S_\psi^\perp$ be its orthogonal complement in~$\mathcal H_D$ ($\dim S_\psi^\perp = D - K$). The required measurement~$M^{(K)}$ exists if $S_\psi^\perp$ is not a completely entangled subspace. This is true for sure, if $\dim S_\psi^\perp > D_E$. Taking into account~(\ref{eq:DimMaxEntSubspace}), after elementary transformations we obtain
\begin{equation}
K \le K_\text{max} = d_1 + \dots + d_l - l. \label{eq:Kmax}
\end{equation}
Therefore if $K \le K_\text{max}$ the factorized estimator-orthogonal measurement~$M^{(K)}$ exists whatever vectors $\ket{\psi_k}$ are, otherwise it may not (which will occur almost certainly in practice).

In numerical simulations and experiments reported here we investigate a bipartite system ($l = 2$) with two identical components, $d_1 = d_2 = \sqrt D$, and $K_\text{max} = 2 \sqrt{D} - 2$. Another notable system is an $l$-qubit register. In this case $K_\text{max} = l$, which is exponentially small in comparison with a maximum possible rank~$2^l$ of a register state.

\subsection{Estimator-orthogonal protocol\label{sec:OrthogonalProtocol}}
There are different ways to incorporate estimator-orthogonal measurements into a particular adaptive protocol,  mainly depending on the system of interest. In the sections below we focus on a bipartite system with two identical parts. The whole system has a Hilbert space $\mathcal H_D = \mathcal H_A \otimes \mathcal H_B$, where~$\mathcal H_A$ and~$\mathcal H_B$ are Hilbert spaces of the subsystems and $\dim \mathcal H_A = \dim \mathcal H_B = \sqrt D$. The following adaptive protocol, which we call an \emph{estimator-orthogonal protocol}, was used in the present work to perform numerical simulations and experiments. It consists of several steps:
\begin{enumerate}
	\item Evaluate the current estimator~$\hat \rho$ of the true state. \label{alg:CalcEstimator}
	\item Calculate the estimator eigenvectors~$\ket{\psi_k}$ and sort them by their corresponding eigenvalues in decreasing order.
	\item Choose index $K$ randomly from the interval $[1, K_\text{max}]$ with a uniform distribution ($K_\text{max} = 2 \sqrt D - 2$).
	\item Find a factorized vector $\ket{\phi_A} \otimes \ket{\phi_B}$, which is simultaneously orthogonal to $K$ eigenvectors: $\scalprod{\phi_A \phi_B}{\psi_k} = 0, k = 1, \dots, K$. \label{alg:FindOrthVec}
	\item Supplement the vector~$\ket{\phi_A}$ with $\sqrt D - 1$ random mutually orthogonal vectors to form a basis~$\mathcal B_A$ in~$\mathcal H_A$. Repeat the analogous procedure for the vector~$\ket{\phi_B}$ to obtain a basis~$\mathcal B_B$. \label{alg:SupplementVec}
	\item Tensorially multiply the basis elements $\ket{\alpha_i} \in \mathcal B_A$ by $\ket{\beta_j} \in \mathcal B_B$ to obtain a basis~$\mathcal B_D$ in $\mathcal H_D$ with elements $\ket{\delta_k} = \ket{\alpha_i} \otimes \ket{\beta_j}$ for all $i, j = 1, \dots, \sqrt D$.
	\item Perform projective measurements in the basis~$\mathcal B_D$. \label{alg:PerformMeas}
	\item Return to the step~\ref{alg:CalcEstimator}, if the total number of registered outcomes~$N$ is less then desired, otherwise stop tomography.
\end{enumerate}

Let us explain some steps in more details. The estimator at the step~\ref{alg:CalcEstimator} is the maximum likelihood estimator. The optimization itself is carried out by means of an accelerated projective gradient (APG) algorithm with adaptive restart~\cite{Candes_FCM2015} (see~\cite{Shang_PRA2017} for a combination of APG with a conjugate gradient method). An initial guess for the APG routine is supplied by the estimator found on the previous iteration of the protocol (on the first iteration a completely mixed state is substituted).

An essential part of the APG method is a projection operation~-- a map of an arbitrary matrix to a space of physical density matrices with a given rank~$R_e$. A common choice is to use a projection, which affects only the eigenvalues of an estimator leaving the eigenvectors unchanged. Therefore, for a full-rank estimate, $R_e = D$, it is sufficient to project a vector of eigenvalues $\lambda \in \mathbb R^D$ onto a canonical simplex $\Delta^{D} = \{\lambda \mid \lambda_i > 0 \wedge \sum_{i = 1}^{D} \lambda_i = 1\}$~\cite{Chen_Arxiv2011}. In the case $R_e < D$, the eigenvalues with indexes $i = R_e + 1, \dots, D$ are zeroed, and a truncated vector~$\lambda \in \mathbb R^{R_e}$ including only the nonzero eigenvalues is subject to projection onto a simplex~$\Delta^{R_e}$.

An orthogonal factorized vector at the step~$\ref{alg:FindOrthVec}$ is found by minimization of the function
\begin{multline}
	f(\ket{\phi_A}, \ket{\phi_B}) = \sum_{k=1}^{K} |\scalprod{\phi_A \phi_B}{\psi_k}|^2 + \\
	\scalprod{\phi_A}{\phi_A} + \frac{1}{\scalprod{\phi_A}{\phi_A}} +
	\scalprod{\phi_B}{\phi_B} + \frac{1}{\scalprod{\phi_B}{\phi_B}} - 4. \label{eq:TargetFucntion}
\end{multline}
This nonnegative function is equal to zero if and only if the vector~$\ket{\phi_A} \otimes \ket{\phi_B}$ is orthogonal to all eigenvectors~$\ket{\psi_k}$ for $k = 1, \dots, K$, and $\ket{\phi_A}$, $\ket{\phi_B}$ are normalized to unit magnitude. Therefore the global minimum $f = 0$ is delivered by the vector being sought. It is guaranteed to exists, because~$K \le K_\text{max}$. Note, that the function~$f$ is constructed to keep the normalization condition and, consequently, the optimization can be accomplished by any unconstrained minimization routine. In particular, we use a Broyden~-- Fletcher~-- Goldfarb~-- Shanno (BFGS) algorithm~\cite{Fletcher_2013}. The algorithm starts from random seed vectors~$\ket{\phi_A}$ and~$\ket{\phi_B}$ having a Haar-uniform distribution~\footnote{\protect{Here ``a vector $\ket{\phi}$ has Haar-uniform distribution'' stands for a jargonized version of ``a random vector~$\ket{\phi}$ has the distribution induced by the Haar measure over the unitary group~$\mathcal U$''. The vector~$\ket{\phi}$ can be expressed as $\ket{\phi} = U \ket{\phi_0}$, where~$U$ is a random unitary matrix, distributed with respect to the Haar measure over~$\mathcal U$, and~$\ket{\phi_0}$ is an arbitrary but fixed vector. Equivalently, if real and imaginary parts of the vector elements are viewed as coordinates of a point in a multidimensional space, this point is uniformly distributed on a sphere.}}. If the algorithm sticks in some local minimum $f \ne 0$, then the minimum is neglected, and optimization is restarted. The global optimum $f = 0$ may be attained for a number of different vectors, and random seed ensures that the algorithm can converge to any of them. This random seed and an overcomplete nature of the protocol provides that the number of various estimator-orthogonal measurements is sufficient.

A random orthonormal basis~$\mathcal B$ containing a given vector~$\ket{\phi}$, required at the step~\ref{alg:SupplementVec}, can be obtained as follows. For a start, note, that any unitary matrix corresponds to some basis and vice versa. A natural measure on a unitary matrix group is a Haar measure, which induces a ``uniform'' distribution on bases. A simple algorithm for the generation of Haar-distributed unitary matrices is known in literature~\cite{Mezzadri_AMS07}. In the beginning, the matrix~$G$, pertaining to the Ginibre ensemble, is taken. By definition, real and imaginary parts of the matrix elements of~$G$ are independent and identically distributed Gaussian random quantities with zero mean and unit variance~\cite{Ginibre65}. Then a QR decomposition is applied to the matrix~$G$, $G = QR$, where~$Q$ is a unitary matrix and~$R$ has a right-triangular form with positive elements on its diagonal. The obtained $Q$ is distributed according to the Haar measure. The vector~$\ket{\phi}$ can be complemented to form a basis~$\mathcal B$ by simply replacing the first column of~$G$ by~$\ket{\phi}$ in the aforementioned procedure. The first column of~$Q$ is also equal to~$\ket{\phi}$ due to the special form of~$R$. Therefore, the matrix~$Q$ corresponds to the basis~$\mathcal B$ being sought.

The basis~$\mathcal B_D$ at the step~\ref{alg:PerformMeas} corresponds to some POVM~$\mathcal M_\alpha$ consisting of rank-1 projectors~$M_{\alpha \gamma}$ onto the basis elements, where an index~$\alpha$ enumerates the bases and~$\gamma$ is an index of an element in the basis. In our experiments the data are collected for a fixed time~$t_\alpha$ for each operator~$M_{\alpha \gamma}$ within the same POVM. Moreover, $t_\alpha$ remains constant for $D+1$ successive bases~-- the minimal number of bases to provide informational completeness for a full-rank estimate. After that~$t_\alpha$ is allowed to change. The change of measurement time~$t_\alpha$ is chosen such that the data \emph{block size} follows some schedule. By the block size we mean an average number of counts $\bigl \langle \sum_{\gamma} n_{\alpha \gamma} \bigr \rangle$ accumulated for a single POVM element. The block size is equal to the likelihood parameter $b_{\alpha \gamma} \equiv b_\alpha$ [see Eq.~(\ref{eq:PoissonLikelihood})]. Previously it was shown that the schedule $b_\alpha \propto N$, where~$N$ is the total number of counts observed so far, is a reasonable trade-off between the benefit from adaptivity and  the computational and measurement realigning overhead~\cite{Kulik_PRA16}. In particular, we use $b_\alpha = \max(100, \lfloor N/30 \rfloor)$ throughout the present work. In the simulations the outcomes are generated using a multinomial likelihood (unlike a Poissonian likelihood in real experiments), so there is no notion of measurement time. The block size is a parameter to be set directly, rather than a quantity depending on measurement time.

\section{Simulations\label{sec:Simulations}}
\subsection{Averaged performance}
We compare the factorized estimator-orthogonal (FO) protocol, described in Sec.~\ref{sec:OrthogonalProtocol}, with four other measurement strategies. All protocols constitute of projective measurements in some informationally overcomplete set of bases. Measurement time and the block size schedule are the same to that of the FO protocol. Protocol abbreviations, used throughout the present work, together with their description are given in the following list:
\begin{enumerate}
	\item FR (factorized random)~-- measurements are performed in random bases consisting of factorized vectors only. The resulting basis is obtained by element-wise tensor product of two subsystem bases, distributed with respect to Haar measure, for every possible pair of their elements.
	\item GR (general random)~-- measurements are performed in random bases of general form drawn from a Haar-uniform distribution.
	\item Eigen~-- an adaptive protocol, which includes measurements in the eigenbasis of the current estimator. The first basis coincides with the eigenbasis, the successive $D$ bases are the GR ones, and they are added to provide informational completeness. When this set of $D+1$ bases is measured, the procedure is repeated: the estimator is updated and a refined eigenbasis is available.
	\item AMUB (aligned mutually unbiased bases)~-- measurements are performed in mutually unbiased bases (MUB)~\cite{Wootters89,Vatan_Alg2002}, rotated in such a way that one of the MUBs coincides with the estimator eigenbasis. Similarly to the Eigen protocol, the procedure is repeated for the successive $D+1$ bases. This protocol is a straightforward extension of an adaptive algorithm proposed in~\cite{Steinberg_PRL13} to high-dimensional systems (the only difference is that MUBs are realigned many times, not only once).
\end{enumerate}
FR and FO protocols utilize only factorized measurements, while others include projectors onto entangled states almost certainly.

We note, that a maximal set of $D+1$ MUBs is known to exist if the dimensionality~$D$ of the system Hilbert space is a power of a prime: $D = p^m$, where $p$ is prime and $m$ is a positive integer. For other dimensions its existence is still an open question. Therefore, the AMUB protocol is not accessible for certain dimensions, and that is why the Eigen protocol is introduced. Actually, these protocols are quite similar, because the ``most important''~-- estimator-orthogonal~-- part, providing the improvement in accuracy, namely, measurements in the eigenbasis is the same for both of them.

The quantitative criterion for the protocol comparison is the Bures distance between the true state~$\rho$ and the estimator~$\hat \rho$:
\begin{equation}
d_B^2(\rho, \hat \rho) = 2 - 2 \sqrt{F(\rho, \hat \rho)} \approx 1 - F(\rho, \hat \rho). \label{eq:BuresMetric}
\end{equation}
The last approximate equality holds in the asymptotic limit $1 - F \ll 1$, therefore, the theory set forth in Sec.~\ref{sec:FidelityDistribution} is also applicable for the squared Bures distance.

\begin{figure*}[tb]
	\centering
	\subfloat[Pure states.]
	{
		\includegraphics[width=0.49\linewidth]{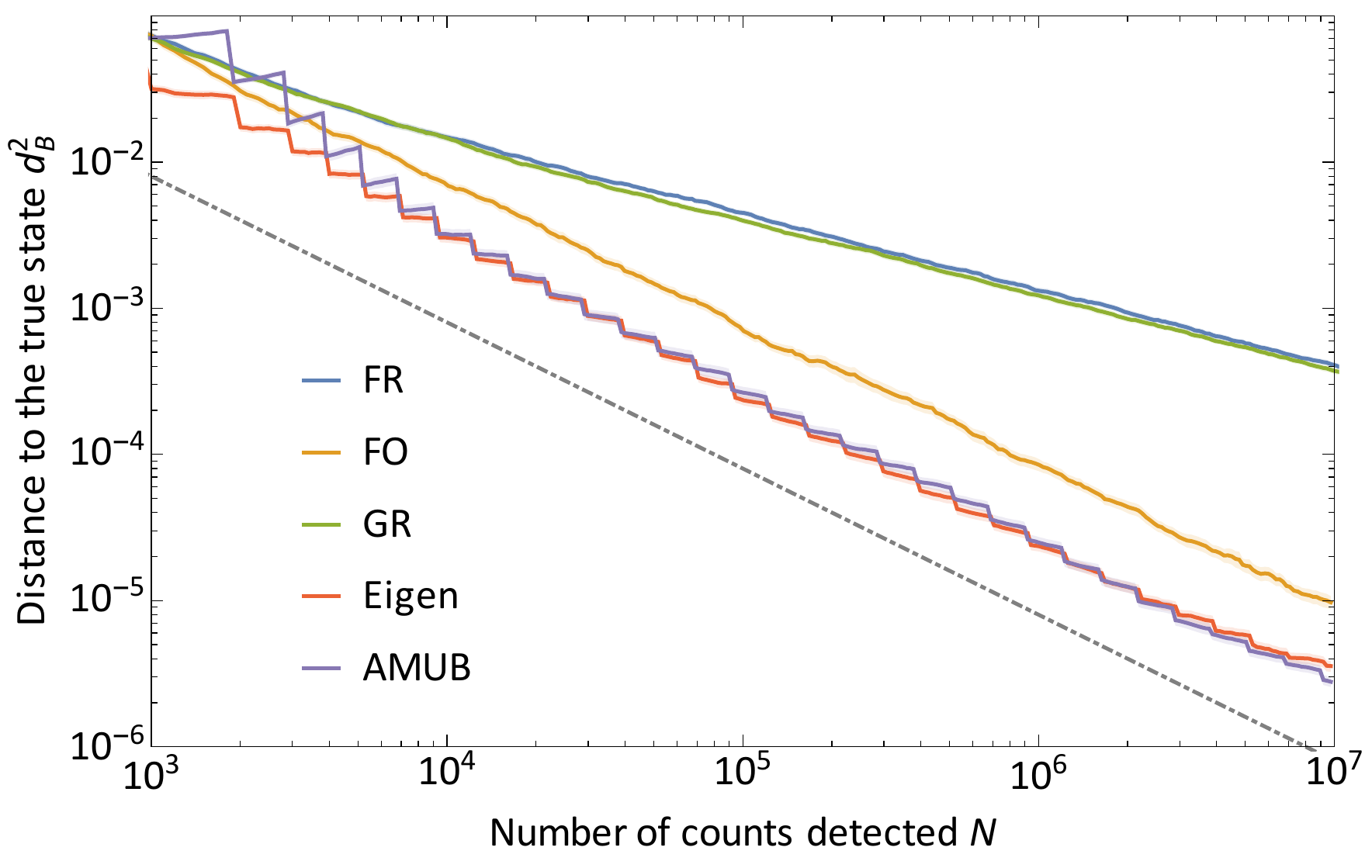}
		\label{fig:PureD9Sim}
	}
	\subfloat[Bures-distributed states.]
	{
		\includegraphics[width=0.49\linewidth]{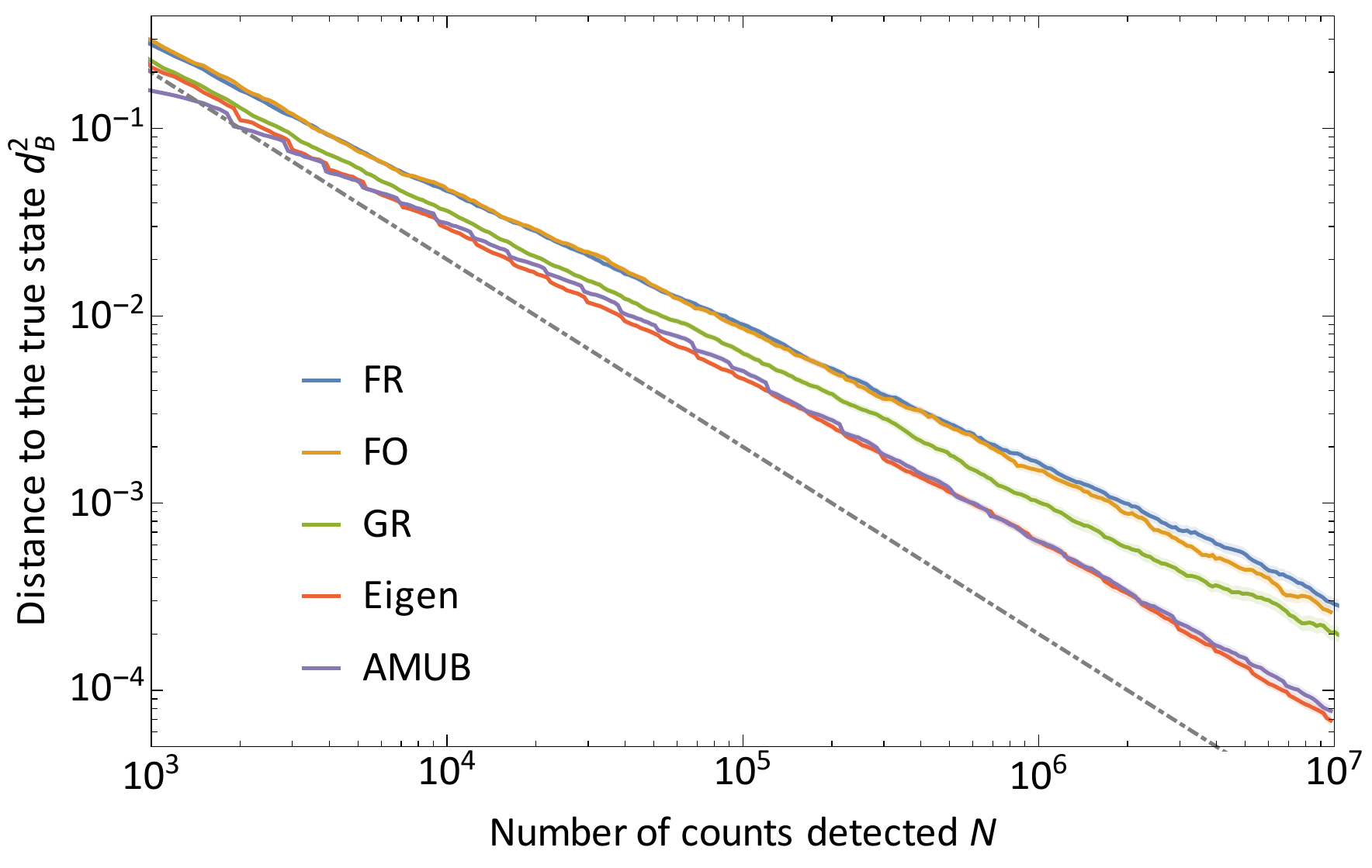}
		\label{fig:BuresD9Sim}
	}
	\caption{The results of numerical simulations for 9-dimensional states. The dependence of the squared Bures distance between the current estimator and the true state on the total number of detected counts~$N$ is shown. Each curve represents performance averaged over Haar-randomly selected pure states (a), and mixed states distributed with respect to the Bures distance-induced measure (b). Here and in subsequent plots FR denotes factorized random measurements, FO~-- factorized estimator-orthogonal protocol, GR~-- random measurements of general form, Eigen~-- measurements in the eigenbasis of the current estimator, AMUB~-- aligned mutually unbiased bases. Dot-dashed lines are Gill~-- Massar bounds $d_B^2 = 8/N$ and $d_B^2 = 200/N$ for pure and mixed state estimation respectively~\cite{Massar_PRA00,Bogdanov_PRA2011}.}
	\label{fig:D9Sim}
\end{figure*}

Dependencies of the Bures distance~$d_B^2(N)$ from the true state~$\rho$ to the current estimator~$\hat \rho(N)$ on the number of counts~$N$ detected are depicted in Fig.~\ref{fig:D9Sim} for a $D = 9$ dimensional system. Unless otherwise is specified, a full-rank estimate with $R_e = D$ is used. The dependencies are averaged over 50 full runs of tomography for different protocols. Two cases are studied: averaged performance among pure Haar-distributed true states (Fig.~\ref{fig:PureD9Sim}) and true states, distributed with respect to measure induced by the Bures metric (Fig.~\ref{fig:BuresD9Sim})~\cite{Osipov10}. All dependencies are well fitted by a power law model $c N^a$. In what follows~$c$ is reffered to as a \emph{prefactor} and~$a$ is called a \emph{convergence rate}. Results of this approximation are summarized in Table~\ref{tab:SimFits}.

\begin{table}[t]
	\caption{Approximation of the dependence of the distance to the true state~$d_B^2(\rho, \hat \rho)$ on the number of counts detected~$N$, obtained in the simulations, with a $c N^a$ model.}
	\label{tab:SimFits}
	\begin{ruledtabular}
			\begin{tabular}{rccll44}
				$D$ & State & Protocol & \multicolumn{1}{c}{$c$} & \multicolumn{1}{c}{$\Delta c$} & \multicolumn{1}{c}{$a$} & \multicolumn{1}{c}{$\Delta a$}\\
				\hline \rule{0pt}{9pt}
				9	& pure	& FR	& 1.73	& 0.07	& -0.519	& 0.003 \\
					&average& FO	& 52	& 4		& -0.967	& 0.006 \\
					& 		& GR	& 1.53	& 0.06	& -0.516	& 0.003 \\
					& 		& Eigen	& 33.8	& 2.2	& -1.019	& 0.005 \\
					& 		& AMUB	& 44	& 3		& -1.038	& 0.006 \\
				\hline \rule{0pt}{9pt}
				9	& Bures	& FR	& 37.9	& 1.4	& -0.728	& 0.003 \\
					&average& FO	& 51.6	& 1.9	& -0.757	& 0.003 \\
					& 		& GR	& 41.4	& 1.6	& -0.765	& 0.003 \\
					& 		& Eigen	& 90.7	& 2.9	& -0.8642	& 0.0028 \\
					& 		& AMUB	& 100	& 3		& -0.8669	& 0.0027 \\
				\hline \rule{0pt}{9pt}
				36	& pure	& FR	& 4.23	& 0.13	& -0.5082	& 0.0020 \\
					&average& FO	& 396	& 30	& -0.874	& 0.005 \\
					& 		& GR	& 4.88	& 0.15	& -0.5159	& 0.0019 \\
					& 		& Eigen	& 131	& 12	& -0.993	& 0.005 \\
			\end{tabular}
	\end{ruledtabular}
\end{table}

For pure states the adaptive strategies are advantageous as they demonstrate $1/N$ convergence, compared to $1/\sqrt{N}$ scaling for the random ones. However, the FO protocol yields to Eigen and AMUB by a prefactor being $\approx 3$ times larger. There is no difference between the FR and GR protocols. As expected, the Eigen and AMUB protocols behave almost similarly (with a slight preponderance of the Eigen protocol for moderate values of~$N$). A characteristic saw-tooth form of~$d_B^2(N)$ dependencies for these protocols is explained by the fact, that it is the measurement in the eigenbasis which significantly refines the current estimator, and the distance to the true state suddenly drops after this measurement is performed.

The dependencies~$d_B^2(N)$ are tighter for different protocols when averaged over Bures-distributed mixed states in contrast to the pure state case. One can still isolate three groups of protocols according to the accuracy they achieve. Strategies including solely factorized measurements have a convergence rate of $a \approx -3/4$ regardless of adaptivity (FR and FO). The GR protocol is better by a prefactor. The most precise protocols with improved convergence rates utilize general type measurements and benefit from adaptivity (Eigen, AMUB).

Previously, nearly the same influence of measurement factorization and adaptivity on tomographic accuracy for pure (mixed) state tomography was observed for a completely different Bayesian approach to state estimation and protocol design~\cite{Kulik_PRA16}.

\begin{figure}[tb]
	\centering
	\includegraphics[width=1.0\linewidth]{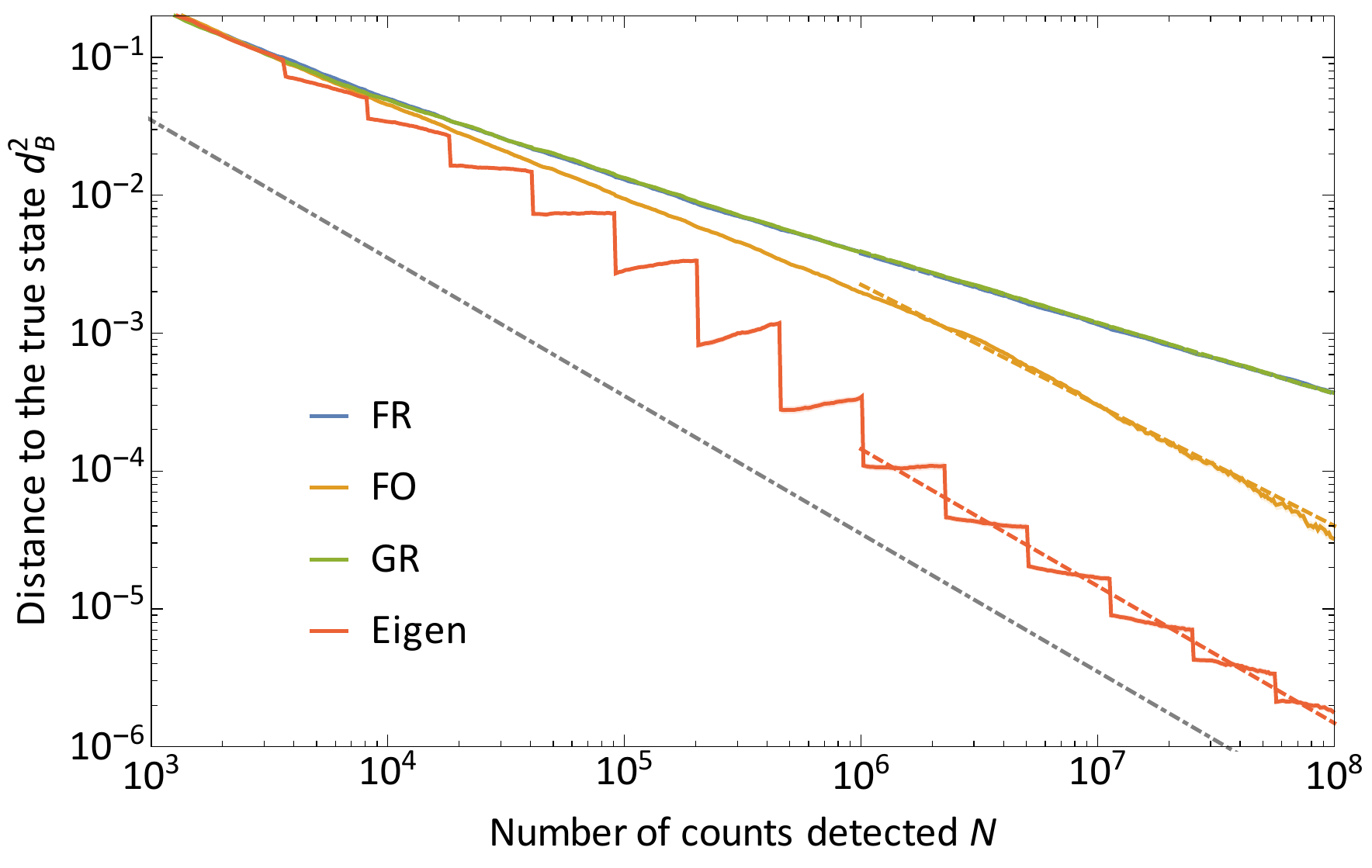}
	\caption{The averaged results of numerical simulations for 36-dimensional random pure true states. Solid curves show the dependence of the squared Bures distance to the true state on the number of registered events~$N$ for different protocols. Dashed lines depict the best fit with the power-law model~$c N^a$. Dot-dashed line is the Gill~-- Massar bound $d_B^2 = 35/N$ for pure state estimation.}
	\label{fig:PureD36Sim}
\end{figure}

We also tested the performance of the aforementioned protocols in pure-state tomography of a 36-dimensional system (with the exception of the AMUB one, which is unavailable for this dimensionality). Again, the dependencies~$d_B^2(N)$ are averaged over 50 full tomography runs for different Haar-distributed true states (see Fig.~\ref{fig:PureD36Sim}). Generally, the results conform to the 9-dimensional case. However, in the case of increased dimensionality the asymptotically optimal convergence rate of the FO protocol is reached for a significantly higher $N$. The transient region of reduced performance seems to increase with growing dimensionality. The parameters of power law fits are presented in Table~\ref{tab:SimFits}.

\subsection{Full vs. adequate-rank estimation}
Convergence of infidelity $1 - F \propto 1/\sqrt{N}$ with the number of counts detected~$N$ occurs only in the situation of rank mismatch, when the true state has lower rank than the estimator, $R_s < R_e$. If the ranks are equal, $R_s = R_e$, then eventually in the asymptotic limit~$N \to \infty$ the $1/N$ convergence appears. Therefore it is important to select an \emph{adequate} rank, which by definition provides $1/N$ convergence whatever the protocol is. The author of the original paper~\cite{Bogdanov_JETP2009} suggests to infer the model rank~$R_e$ from the observed data itself using some kind of a $\chi^2$-consistency test.

Formally, all states have full rank in real experiments (but some eigenvalues may be relatively small), and it looks like the full-rank estimation should always be used. But the estimator with the adequate (and partial) rank captures nonzero eigenvalues, which are statistically significant, treating possibly small eigenvalues as essentially zero ones, without any influence on accuracy. Accordingly, an adequate rank can be selected on the fly, while tomography proceeds and new eigenvalues become significant, and in principle this scenario ensures~$1/N$ convergence. It seems that adaptive tomography (and the estimator-orthogonal protocol in particular) can offer nothing more than the adequate rank selection does. However, it appears, that rank selection and adaptivity do not exclude each other~-- one can benefit from both of them.

\begin{figure*}[tb]
	\centering
	\subfloat[Full rank.]
	{
		\includegraphics[width=0.49\linewidth]{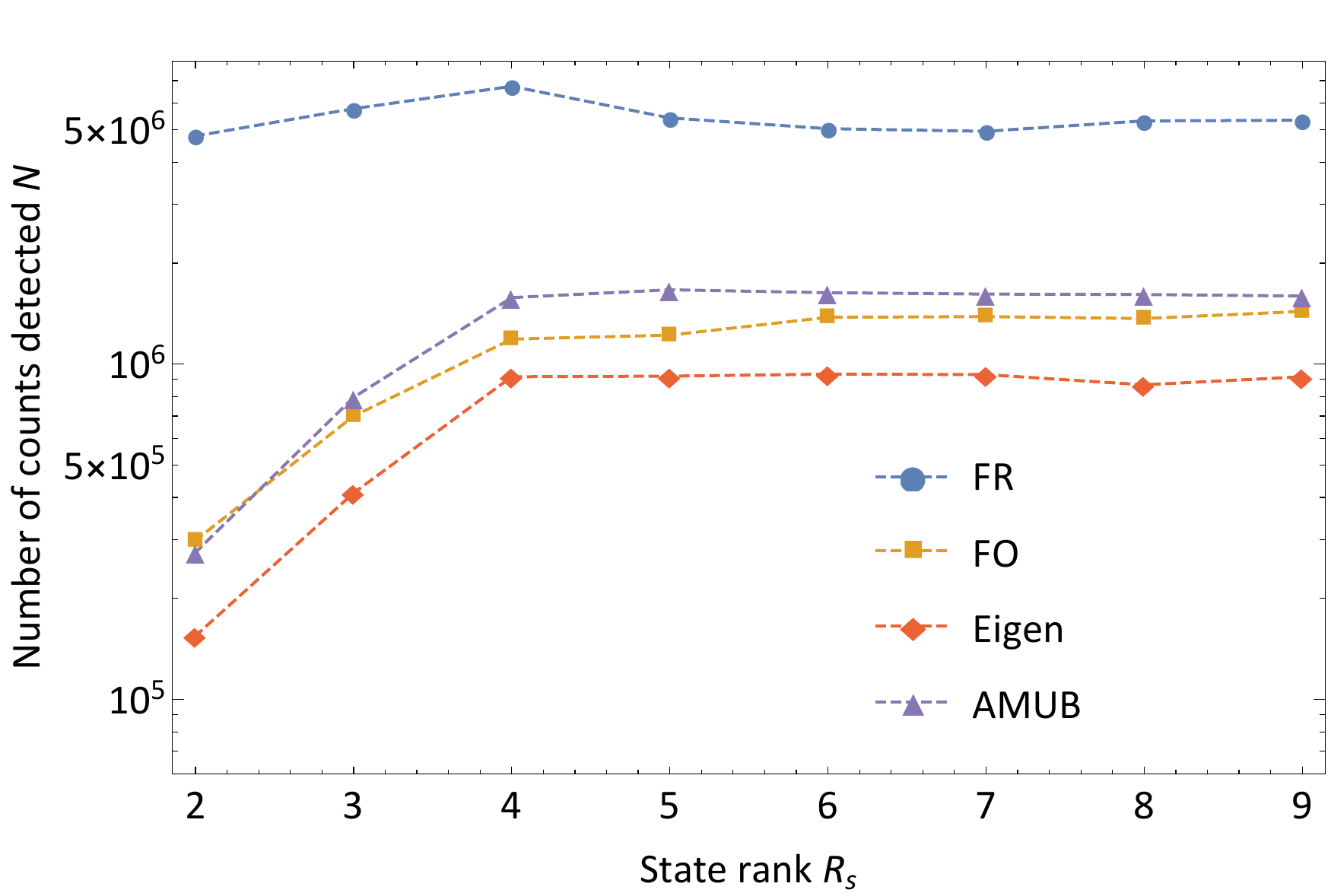}
		\label{fig:ReachNDiffRanks}
	}
	\subfloat[Adequate rank.]
	{
		\includegraphics[width=0.49\linewidth]{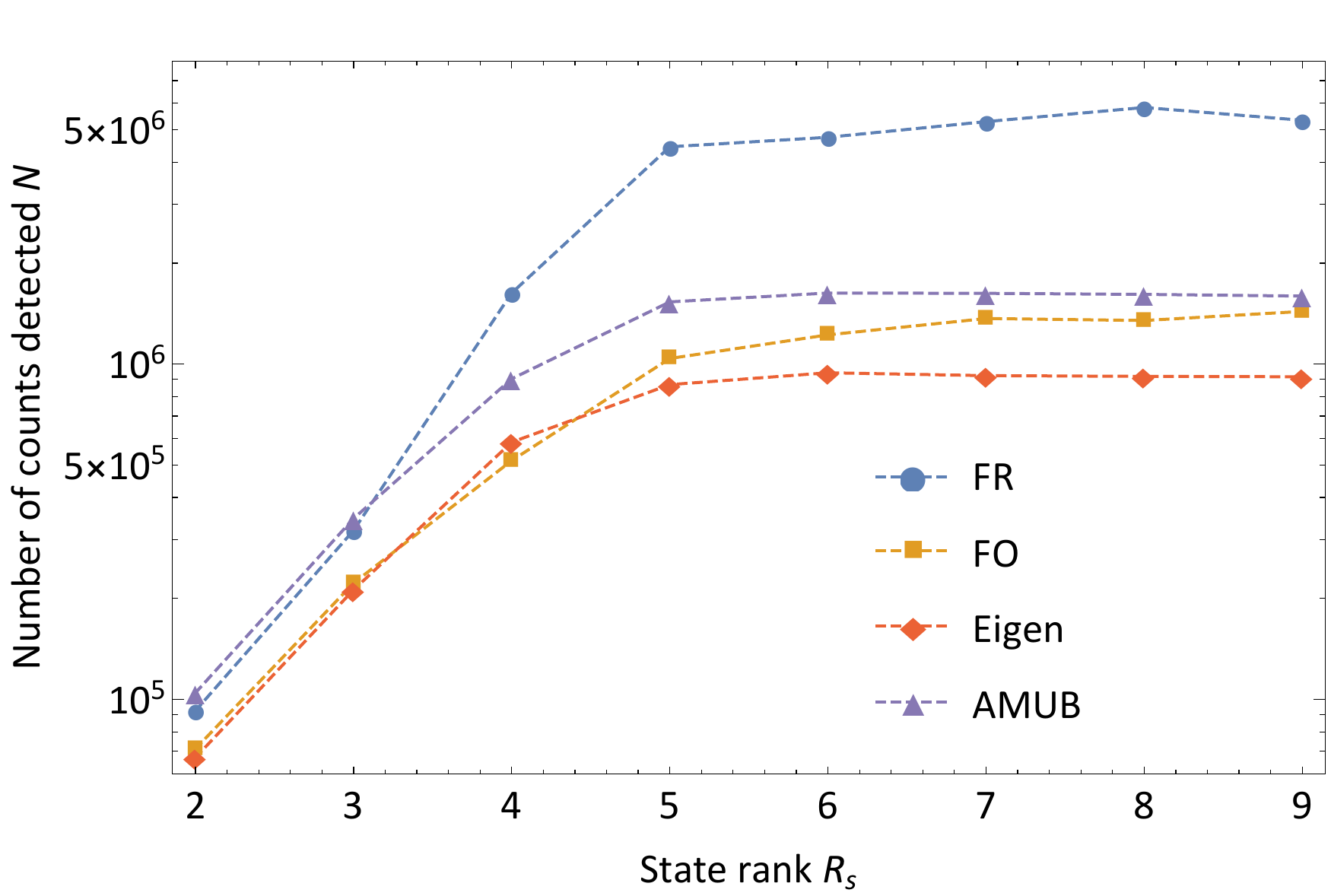}
		\label{fig:ReachNEqualRanks}
	}
	\caption{Change in the total number of detected counts~$N$, required to reach the certain distance $d_B^2(N) = 10^{-3}$ to the true state, with the rank of the true state. Simulated results of full-rank (a) and adequate-rank (b) estimation are shown for different protocols. Dashed lines are guides to the eye.}
	\label{fig:ReachN}
\end{figure*}

We have carried out numerical simulations for the true states with different ranks to reveal the relation between rank selection and adaptivity. Two cases were studied: a full-rank estimate ($R_e = D$) and an optimal-rank inference ($R_e = R_s$). The plots in Fig.~\ref{fig:ReachN} depict the values of~$N$, required to reach a certain value of accuracy $d_B^2(N) = 10^{-3}$, versus the true state rank for different protocols. The underlying dependencies~$d_B^2(N)$ have been averaged over 50 runs of tomography, the utilized true states are inspired by our experimental implementation and are listed in the \appendixname.

When state and estimator ranks disagree (Fig.~\ref{fig:ReachNDiffRanks}) random measurements demonstrate evenly poor performance, as expected, while adaptive protocols are beneficial, especially for low-rank states. FO protocol requires~$\approx 3$ times less amount of statistics~$N$ than FR to achieve the given level of accuracy for ranks $R_s \ge 4$. This advantage increases up to~$\approx 30$ times towards low-rank states. Remarkably, AMUB is slightly less accurate than FO for these particular true states in contrast to the averaged performance (see Fig.~\ref{fig:D9Sim}), even though it uses measurements of general type.

Performance of random measurements changes qualitatively in the situation $R_e = R_s$ (Fig.~\ref{fig:ReachNEqualRanks}). Asymptotically all protocols have~$1/N$ convergence and differ only by prefactors. When this asymptotic becomes valid, random measurements are almost as good as adaptive protocols. It happens for low-rank states, in our case for $R_s \le 4$. However, optimal rank selection has little impact on tomographic accuracy for higher-rank states, and adaptivity provides much more advantage. It is worth to mention, that the particular crossover point $R_s = 4$ depends on the given level of accuracy $d_B^2 = 10^{-3}$ and the true states being simulated. 

\section{Experiment\label{sec:Experiment}}
\begin{figure}[tbh]
	\centering
	\includegraphics[width=1.0\linewidth]{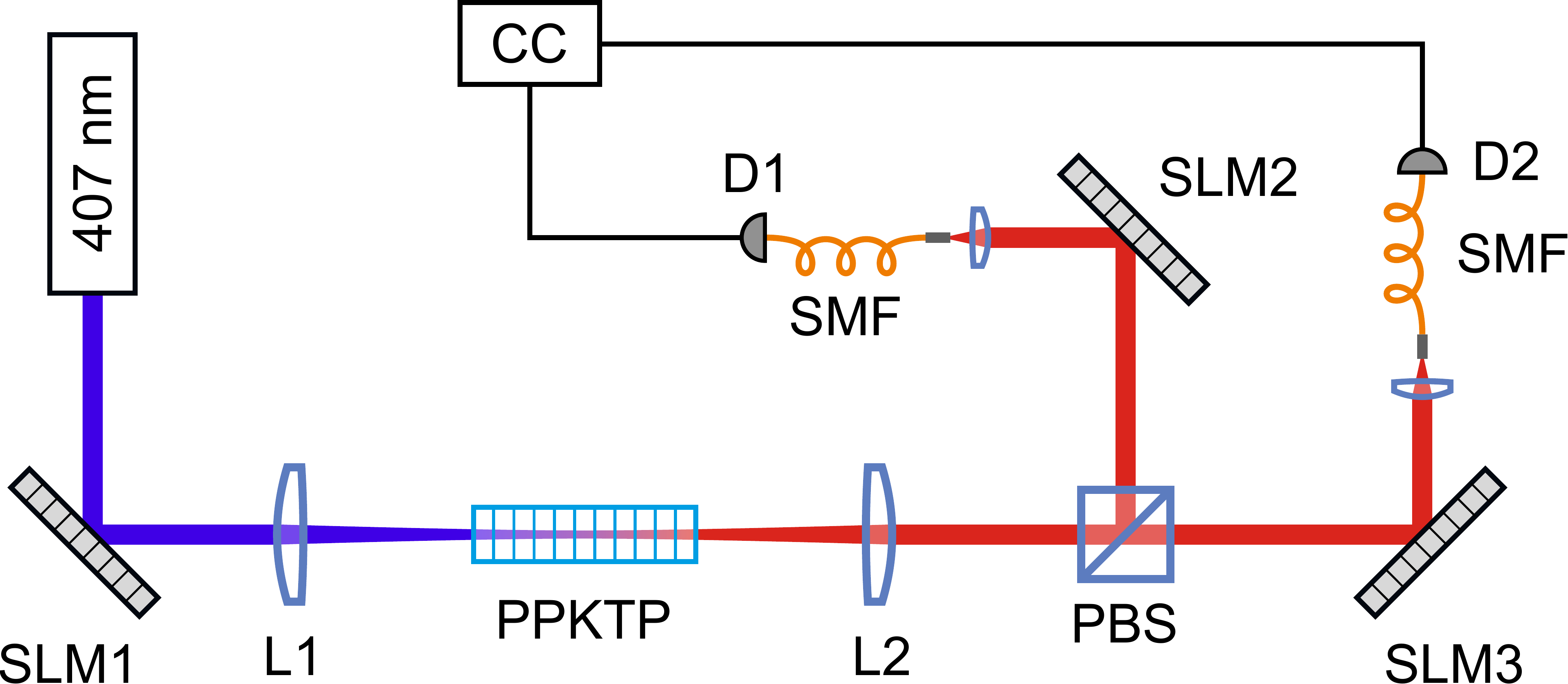}
	\caption{A simplified scheme of the experimental setup. A spatial state of photon pairs, produced in spontaneous parametric down conversion, is controlled by a spatial light modulator SLM1. SLM2 and SLM3 together with single-mode fibers SMF define a projective factorized measurement in the basis of orthogonal spatial modes, performed on a photon pair.}
	\label{fig:Setup}
\end{figure}


The experiment is implemented using spatial degrees of freedom of biphotons generated in spontaneous parametric down conversion (SPDC). We use a conventional measurement scheme consisting of a Hanbury Brown~-- Twiss interferometer equipped with spatial light modulators (SLM) in each arm. A simplified scheme of the setup is shown in Fig.~\ref{fig:Setup} (see Ref.~\cite{Kulik_PRL2017} for a detailed discussion of the experimental setup). Radiation of a 407-nm diode laser, spatially filtered by a single-mode fiber (not shown), is directed onto an SLM1 to form the desired transverse profile of the beam, diffracted into the first order. This beam is served as a pump for a 25-mm-thick periodically poled KTP crystal (PPKTP), designed for a collinear degenerate type-II phase matching. A lens L1 provides an optimal focusing of the pump into the crystal to achieve a single-mode SPDC regime~\cite{Kulik_PRL2017}, while a lens L2 collimates the down-converted radiation. A photon pair is separated into two arms by a polarization beam splitter PBS. SLM2 and SLM3 (actually these are two halves of the same SLM) realize a given transformation of photon spatial states in the first order of diffraction. The diffracted light is collected into single-mode fibers SMF, which perform a projection onto a fundamental (Gaussian) spatial mode. The fibers are connected to single-photon counting modules D1 and D2 followed by a home-made coincidence circuit CC with a 4-ns time window.


A digital hologram displayed on the SLM1 controls the spatial mode of the produced photon pairs, while SLM2-3 holograms together with the SMFs determine a projective measurement. The utilized SLMs are of phase-only nature, but there exists a method to perform amplitude modulation with phase-only holograms as well (that is one of the reasons, why the first diffraction order is used)~\cite{Boyd_OptLett2013}. The experimental setup permits only factorized measurements, because each photon from the pair is directed onto its own SLM and propagates separately.

There are two canonical choices of basis modes: Laguerre~-- Gaussian (LG) and Hermite~-- Gaussian (HG) ones. The privileged role of these modes is based on the fact, that they are eigen solutions of a paraxial wave equation, therefore their shape is preserved during propagation. Moreover, they form full infinite-dimensional orthogonal bases in the space transversal modes. We have chosen HG modes to deal with in our experiments. Field amplitude of the HG mode at the beam waist is given by
\begin{equation}
\HG_{nm}(x, y) \propto H_n \left(\frac{x}{w}\right) H_m \left(\frac{y}{w}\right) \exp \Bigl(-\frac{x^2 + y^2}{2 w^2} \Bigr). \label{eq:HGMode}
\end{equation}
where $n, m$ are nonnegative mode indexes, $x, y$ are transversal coordinates, $w$ is a waist parameter, and $H_n$ is an $n$-th order Hermite polynomial. The \emph{order of a mode} is defined as a sum $n + m$.

We have experimentally prepared two states, a factorized and an entangled one, which approximately correspond to $\ket{\HG_{00}} \otimes \ket{\HG_{00}}$ (Gaussian) and $(\ket{\HG_{10}} \otimes \ket{\HG_{00}} + \ket{\HG_{00}} \otimes \ket{\HG_{10}})/\sqrt{2}$ (Bell). They can be produced by pumping the crystal with, respectively,~$\HG_{00}$ and~$\HG_{10}$ modes with a waist conforming to crystal parameters (a waist of detection modes is also uniquely determined)~\cite{Walborn_JPB2012}. Additionally, the Gaussian state was spatially filtered with a single-mode fiber, installed between the crystal and the PBS (not shown in Fig.~\ref{fig:Setup}), to increase its purity.

Since we are interested only in finite-dimensional tomography, we should limit the dimensionality~$D$ by selecting a certain subspace. The first one we used is the 9-dimensional subspace, spanned by all possible pairwise tensor products of $\ket{\HG_{00}}, \ket{\HG_{01}}, \ket{\HG_{10}}$ modes, e.g. $\ket{\HG_{00}} \otimes \ket{\HG_{01}}$, etc. By appending second-order modes, namely $\ket{\HG_{11}}, \ket{\HG_{20}}, \ket{\HG_{02}}$, another subspace with increased dimensionality $D = 36$ is constructed.

The prepared states can be reconstructed in either subspace. This gives us four combinations, however, in preliminary experiments we found that FO and FR protocols perform equally for the Bell state, reconstructed in a 36-dimensional subspace, therefore this case is excluded from further comparison. We attribute this behaviour to the low purity of the experimentally prepared state. The parameters of the states for the remaining three series of experiments are listed in Table~\ref{tab:StateParameters}. They include purity $\Tr \rho^2$, negativity~\cite{Vidal_PRA2002}, and \emph{spread}~$d_\text{spr}^2$, averaged over several tomography runs (30 runs for $D = 9$ and 10~-- for $D = 36$). The total number of photon pairs detected in each run is $N = 3 \times 10^5$. The spread~$d_\text{spr}^2$ is defined as the averaged Bures distance from each state~$\rho_i$ in the ensemble to the mean state $\bar \rho = \frac1m \sum_{i = 1}^{m} \rho_i$: $d_\text{spr}^2 = \frac1m \sum_{i = 1}^{m} d_B^2(\rho_i, \bar \rho)$. It captures both the statistical uncertainty of the estimator and the systematic drift of the true state from run to run. Our analysis shows that the contribution from the latter prevails. Fluctuations of the true state mainly account for slow variation of the environment temperature, and, besides that, not all runs were contiguous, they were split into several days with some interruption for setup adjustment.

\begin{table}[t]
	\caption{Parameters of the states obtained in the experiment averaged over several tomography runs. See text for parameter definitions.}
	\label{tab:StateParameters}
	\begin{ruledtabular}
		\begin{tabular}{rc D{s}{\,\pm\,}{4.4} D{s}{\,\pm\,}{4.4} D{s}{\,\pm\,}{5.5}}
			$D$ & State & \multicolumn{1}{c}{Purity} & \multicolumn{1}{c}{Negativity} & \multicolumn{1}{c}{Spread}\\
			\hline \rule{0pt}{9pt}
			9	& Gaussian	& 0.942	s 0.015 & 0.003 s 0.001 & 0.0052 s 0.0004 \\
			36	& Gaussian	& 0.915	s 0.007 & 0.022 s 0.002 & 0.062  s 0.013  \\
			9	& Bell	& 0.740	s 0.003 & 0.376 s 0.005 & 0.0168 s 0.0013 \\
		\end{tabular}
	\end{ruledtabular}
\end{table}

\begin{figure*}[tb]
	\centering
	\includegraphics[width=0.99\linewidth]{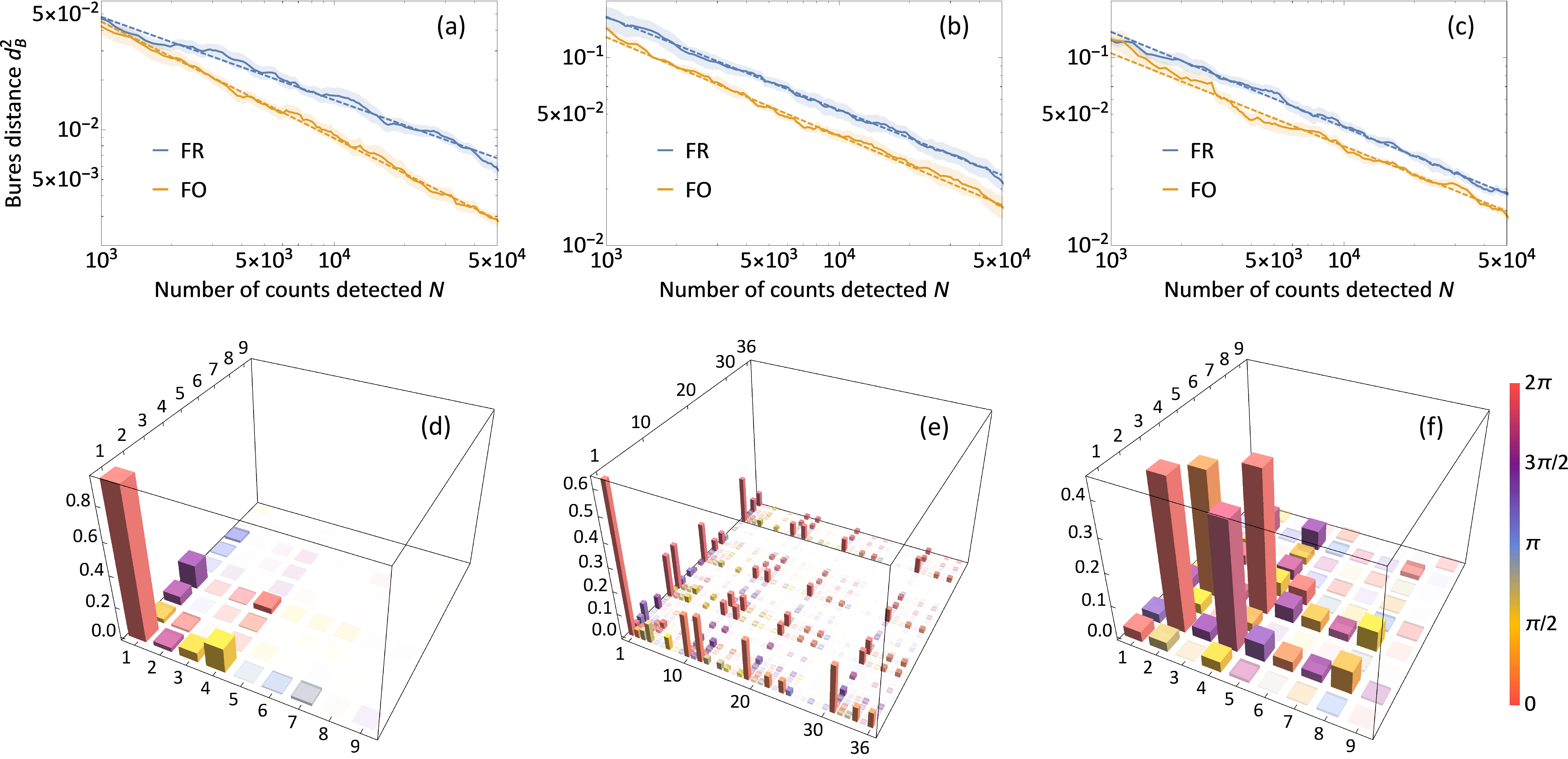}
	\caption{Experimental dependencies of the squared Bures distance to the final estimator on the number of detected photon pairs~$N$ for the Gaussian state, reconstructed in the subspace with dimensionality $D = 9$ (a) and $D = 36$ (b), and for the Bell state, reconstructed in the subspace with dimensionality $D = 9$ (c). The shaded area corresponds to one standard deviation of mean. Dashed lines are power-law fits to the data. Density matrix plots (d-f) of the final estimators are shown under the respective convergence plots. An absolute value of a matrix element corresponds to a bin height, while its phase is encoded by color.}
	\label{fig:Experiment}
	
\end{figure*}

Previously in Sec.~\ref{sec:Simulations} we quantified the accuracy of estimation by the Bures distance~$d_B^2(\hat \rho(N), \rho)$ between the current estimator~$\hat \rho(N)$ and the true state~$\rho$. In the experiment the exact true state is unknown and tomography provides the best estimation at hand. Thus, we resort to the Bures distance~$d_B^2(N) \equiv d_B^2(\hat \rho(N), \hat \rho(N_0))$ to the \emph{final} estimate $\hat \rho(N_0)$, calculated after all the data~$N_0$ is gathered. These dependencies are shown in Fig.~\ref{fig:Experiment}a-\ref{fig:Experiment}c for a full-rank tomography of the Gaussian state and the Bell one. The results are averaged over several tomography runs (from 5 to 20), and the total number of observed counts in each run is $N_0 = 3 \times 10^5$. Obviously, $d_B^2(N)$ tends to be exactly zero, when~$N$ approaches~$N_0$, $d_B^2(N_0) = 0$, therefore the plots are truncated at $N = 5 \times 10^4$ to remove the spurious region. Again, we approximate the dependencies with a power-law model~$c N^a$ (see Table~\ref{tab:ExpFits} for the best-fit parameters). The corresponding density matrix plots of the final estimators are shown in Fig.~\ref{fig:Experiment}d-\ref{fig:Experiment}f.

\begin{table}[h]
	\caption{Approximation of the dependence of the distance to the final estimator~$d_B^2(\hat \rho, \hat \rho(N_0))$ on the number of counts detected~$N$, obtained in experiments, with $c N^a$ model.}
	\label{tab:ExpFits}
	\begin{ruledtabular}
		\begin{tabular}{rcc2233}
			$D$ & State & Protocol & \multicolumn{1}{c}{$c$} & \multicolumn{1}{c}{$\Delta c$} & \multicolumn{1}{c}{$a$} & \multicolumn{1}{c}{$\Delta a$}\\
			\hline \rule{0pt}{9pt}
			9	& Gaussian	& FR	& 1.54	& 0.19	& -0.502	& 0.014 \\
			& 		& FO	& 5.8	& 0.9	& -0.703	& 0.016 \\
			\hline \rule{0pt}{9pt}
			36	& Gaussian	& FR	& 5.1	& 0.7	& -0.496	& 0.016 \\
			& 		& FO	& 4.9	& 0.4	& -0.526	& 0.010 \\
			\hline \rule{0pt}{9pt}
			9	& Bell	& FR	& 4.6	& 0.4	& -0.507	& 0.010 \\
			& 		& FO	& 3.2	& 0.4	& -0.495	& 0.012 \\
		\end{tabular}
	\end{ruledtabular}
\end{table}

One can see, that the FO protocol demonstrates an advantage over random measurements in all considered situations. However, the relative benefit varies, depending on the purity and dimensionality of the true state. The maximal gain occurs for nearly pure state with lower dimensionality (Gaussian, $D = 9$). The FR protocol converges as $1/\sqrt{N}$, while the FO one has an improved convergence rate $a = -0.70$ (Fig.~\ref{fig:Experiment}a). This difference results in $\approx 2.2$ times more accurate estimation for $N = 5 \times 10^4$ (the ratio can be even larger if one would collect larger total statistics~$N_0$). When the Gaussian state is reconstructed in a subspace with higher dimensionality $D = 36$, the convergence rate of the FO protocol becomes nearly the same as for the FR strategy (Fig.~\ref{fig:Experiment}b). However, a steady accuracy improvement of $\approx 1.4$ times is observed. The Bell state with relatively low purity is the ``hardest'' one to estimate. The averaged gap between the Bures-distance dependencies is $\approx 1.25$ times (Fig.~\ref{fig:Experiment}c).

\vfill
	
\section{Conclusion\label{sec:Conclusion}}

\begin{table*}
	\caption{Eigenvalues $\lambda_1, \dots, \lambda_9$ of the true states having different rank~$R_s$, utilized in simulations for plotting Fig.~\ref{fig:ReachN}.}
	\label{tab:Eigenvalues}
	\begin{ruledtabular}
		\begin{tabular}{r|ccccccccc}
			$R_s $ & $ \lambda_1 $ & $ \lambda_2 $ & $ \lambda_3 $ & $ \lambda_4 $ & $ \lambda_5 $ & $ \lambda_6 $ & $ \lambda_7 $ & $ \lambda_8 $ & $ \lambda_9 $ \\
			\hline \rule{0pt}{9pt}
			$2$ & $9.4721\e{-1}$ & $5.2786\e{-2}$ & $0$            & $0$            & $0$            & $0$            & $0$            & $0$            & $0$ \\
			$3$ & $9.4792\e{-1}$ & $3.2883\e{-2}$ & $1.9198\e{-2}$ & $0$            & $0$            & $0$            & $0$            & $0$            & $0$ \\
			$4$ & $9.4798\e{-1}$ & $3.1398\e{-2}$ & $1.8331\e{-2}$ & $2.2868\e{-3}$ & $0$            & $0$            & $0$            & $0$            & $0$ \\
			$5$ & $9.4799\e{-1}$ & $3.1209\e{-2}$ & $1.8221\e{-2}$ & $2.2730\e{-3}$ & $3.0420\e{-4}$ & $0$            & $0$            & $0$            & $0$ \\
			$6$ & $9.4800\e{-1}$ & $3.1094\e{-2}$ & $1.8153\e{-2}$ & $2.2646\e{-3}$ & $3.0308\e{-4}$ & $1.8683\e{-4}$ & $0 $           & $0$            & $0$ \\
			$7$ & $9.4800\e{-1}$ & $3.1061\e{-2}$ & $1.8134\e{-2}$ & $2.2622\e{-3}$ & $3.0276\e{-4}$ & $1.8663\e{-4}$ & $5.3916\e{-5}$ & $0$            & $0$ \\
			$8$ & $9.4800\e{-1}$ & $3.1050\e{-2}$ & $1.8128\e{-2}$ & $2.2614\e{-3}$ & $3.0265\e{-4}$ & $1.8656\e{-4}$ & $5.3896\e{-5}$ & $1.8656\e{-5}$ & $0$ \\
			$9$ & $9.4800\e{-1}$ & $3.1045\e{-2}$ & $1.8125\e{-2}$ & $2.2611\e{-3}$ & $3.0261\e{-4}$ & $1.8654\e{-4}$ & $5.3889\e{-5}$ & $1.8654\e{-5}$ & $7.4615\e{-6}$ \\
		\end{tabular}
	\end{ruledtabular}
\end{table*}

We have experimentally demonstrated the feasibility of obtaining advantage in the reconstruction infidelity for an adaptive tomography protocol for the states, living in the Hilbert space of dimensionality as high as $D=36$. The main innovation here is a simple adaptive protocol specially tailored for factorized measurements, and thus ideally suited for bipartite systems, such as SPDC photon pairs used in the experiment. The protocol is completely agnostic to the origin of the estimation procedure, i.e. it requires only the point estimate of the state density matrix. So this adaptive optimization may supplement any tomographic procedure, both Bayesian and frequentist in nature. It is practically attractive, because the optimization routine involved in the search of optimal measurements is very fast. In this respect it may be considered as a generalization of the two-step strategy used in \cite{Steinberg_PRL13} to high-dimensional systems. This generalization, however, explicitly avoids entangled projectors, thus making it experimentally feasible.

Since there is almost no additional overhead for adaptive optimization, the protocol may be used and provide advantage whenever the state estimation itself is feasible. This boundary is unfortunately not that far from the dimensionality of the system used in this work. To the best of our knowledge, the current record for full-tomography is a 14 qubit simulation performed in \cite{Hou_NJP2016} which took 4 hours of computational time. It is hardly possible to extend the full reconstruction much further. Therefore methods are developed to trade the completeness of reconstruction for efficiency. For example, one may utilize some properties of the state known a priory, like sparseness (low rank) of the density matrix \cite{Eisert_PRL10} or its tensor-product structure \cite{Plenio_NatureComm10,Plenio_PRL2013}. Whether such scalable protocols, providing partial information about the state, may enjoy the advantage from adaptivity is an interesting open question. Another option for further work is the generalization of the protocol to process tomography. Although the Choi~-- Jamio\l kowski isomorphism formally reduces process tomography to state tomography, additional restrictions on which probe states and measurements may be realized in experiment pose additional constraints, which should be carefully treated. For example, a standard prepare-and-measure scenario in process tomography corresponds to factorized measurements on the Choi~-- Jamio\l kowski state, making the estimator-orthogonal protocol discussed here a natural choice. These questions will be addressed elsewhere.

\begin{acknowledgments}
This work was funded by the Russian Science Foundation grant \#~16-12-00017. GIS and EVK acknowledge the support of Foundation for the advancement of theoretical physics and mathematics ``BASIS''. 
\end{acknowledgments}

\vfill

\appendix*
\section{List of true states used in simulations\label{sec:TrueStates}}
In this Appendix we present a list of the true states of different rank~$R_s$, utilized to obtain the data for Fig.~\ref{fig:ReachN}. As a fiducial state we have taken one of Bell states, recovered in the experiment. This state possesses a full rank and has purity of $\approx 0.74$. Its smallest eigenvalues are zeroed to derive the states with smaller ranks. After normalization to unit trace, the purity of the states is artificially set to be equal to $0.90$ by increasing the weight of the first eigenvector~$\ket{\psi_1}$: a state~$\rho$ is replaced by $(1 - \mu) \rho + \mu \dyad{\psi_1}$ with an appropriately chosen coefficient~$\mu$. This procedure leads to states of the form $\rho_{R_s} = U \Lambda_{R_s} U^\dagger$ with the same matrix of eigenvectors~$U$ and different diagonal matrices of eigenvalues $\Lambda_{R_s} = \diag(\lambda_1, \dots, \lambda_9)$. The corresponding eigenvalues are listed in Table~\ref{tab:Eigenvalues}, and the matrix~$U$ reads,
\vfill
\begin{widetext}
\begin{multline}
U = 
\left[
\begin{array}{555555555}
 0.18499 &  0.33521 &  0.39189 &  0.11521 &  0.70821 &  0.16251 &  0.16181 &  0.10863 &  0.34728 \\
 0.28549 &  0.04306 & -0.12139 & -0.09399 & -0.02245 & -0.07123 &  0.10103 &  0.04967 & -0.00898 \\
 0.08197 &  0.13167 &  0.07218 & -0.02060 &  0.05490 & -0.12759 & -0.20933 & -0.41788 & -0.06937 \\
-0.36676 &  0.23376 & -0.10614 & -0.01045 &  0.00661 & -0.12757 & -0.04444 &  0.04407 &  0.14611 \\
-0.14044 &  0.34180 &  0.61698 &  0.02377 & -0.34047 & -0.29427 &  0.17399 &  0.10504 & -0.24112 \\
 0.02838 &  0.02135 &  0.26563 &  0.26694 & -0.14372 & -0.40065 & -0.18730 &  0.10913 &  0.10968 \\
 0.06681 &  0.04542 &  0.09029 & -0.02554 & -0.28469 &  0.41781 & -0.34626 &  0.19022 &  0.31406 \\
-0.12165 & -0.24829 &  0.10246 &  0.04349 & -0.32817 &  0.00200 &  0.33902 & -0.01463 &  0.68935 \\
-0.02202 &  0.05324 & -0.21750 &  0.45973 &  0.01965 & -0.22045 &  0.21995 & -0.57611 &  0.19407 \\
\end{array}
\right] \\
+ i \left[
\begin{array}{555555555}
 0       &  0       &  0       &  0       &  0       &  0       &  0       &  0       &  0       \\
-0.63903 & -0.50160 &  0.28092 & -0.03901 &  0.32871 &  0.01572 & -0.10512 & -0.09365 & -0.07893 \\
 0.00234 &  0.21111 &  0.10821 & -0.49839 &  0.00305 & -0.08849 & -0.38590 & -0.47498 &  0.20179 \\
 0.52400 & -0.56738 &  0.35821 & -0.00782 &  0.01201 & -0.09150 & -0.05189 & -0.14004 & -0.04678 \\
-0.09085 &  0.03645 &  0.05253 & -0.03338 & -0.04537 &  0.30683 &  0.18784 & -0.13416 & -0.13161 \\
-0.01911 & -0.06071 & -0.19854 &  0.22619 &  0.16048 & -0.35704 & -0.48028 &  0.32552 &  0.17954 \\
-0.05090 &  0.06307 &  0.19114 &  0.31363 & -0.17868 &  0.37403 & -0.35105 & -0.18574 &  0.05752 \\
-0.07944 & -0.00239 & -0.01302 & -0.39594 & -0.01530 & -0.04769 &  0.12546 &  0.03970 &  0.17332 \\
-0.04804 &  0.05369 &  0.01458 &  0.36643 & -0.03133 &  0.29230 & -0.04073 & -0.06245 & -0.19862 \\
\end{array}
\right]. \label{eq:Eigenvectors}
\end{multline}
\end{widetext}

\bibliographystyle{apsrev4-1}
\bibliography{ref_base}

\end{document}